Analysis of the Benefits and Efficacy of the Addition of Variants and Reality Paths to the Blackboard Architecture

Ben Clark, Matthew Tassava, Cameron Kolodjski & Jeremy Straub
Institute for Cyber Security Education and Research
North Dakota State University
1320 Albrecht Blvd., Room 258
Fargo, ND 58102
Phone: +1 (701) 231-8196
Fax: +1 (701) 231-8255
Email: benjamin.clark.3@ndsu.edu, matthew.tassava@ndsu.edu, cameron.kolodjski@ndsu.edu, jeremy.straub@ndsu.edu

**Abstract**

While the Blackboard Architecture has been in use since the 1980s, it has recently been proposed for modeling computer networks to assess their security. To do this, it must account for complex network attack patterns involving multiple attack routes and possible mid-attack system state changes. This paper proposes a data structure which can be used to model paths from an ingress point to a given egress point in Blackboard Architecture-modeled computer networks. It is designed to contain the pertinent information required for a systematic traversal through a changing network. This structure, called a reality path, represents a single potential pathway through the network with a given set of facts in a particular sequence of states. Another structure, called variants, is used during traversal of nodes (called containers) modeled in the network. The two structures – reality paths and variants – facilitate the use of a traversal algorithm, which will find all possible attack paths in Blackboard Architecture-modeled networks. This paper introduces and assesses the efficacy of variants and reality paths

## 1. Introduction

Numerous search algorithms, such as depth-first searches and iterative deepening searches, perform suitably when tasked with searching for a single, efficient path to a goal through a simple state space. However, some applications require searches with alternate goals. One such application is security vulnerability identification. While an attacker might seek to find an optimal path through a network of systems to reach their attack target, defenders need to identify all possible attack paths to facilitate their elimination or mitigation.

A system for identifying vulnerability paths was previously proposed which utilized the Blackboard Architecture and introduced links and containers to model computer networks' physical architecture, leaving rules, facts and actions to model logical interactions. During the operations of this system, fact values inherently change (in some cases, multiple times) as operations occur within the Blackboard Architecture network simulating the real-world computer network. In this system, vulnerability exploitation is represented by rules, which can be run if triggered by their preconditions being satisfied. As some facts can be toggled back and forth between true and false during operations, it is useful to understand when and how certain fact values (and combinations of fact values) can be reached to satisfy rule pre-conditions.

Because of this, for the purposes of the proposed system, simple traversal algorithms will not work for several reasons. Most significantly, although the proposed system, like typical traversal algorithms, aims to find paths from an ingress point in a network to an egress point, it specifically aims to find all possible attack paths in a network. Finding only a single path or a subset of the set of possible paths would not achieve the goal of trying to mitigate all possible paths of attack. Additionally, while exhaustive traversal algorithms exist for simple networks, it is possible for rules to change facts and, thus, enable new paths of traversal at any time. These paths may use containers which have previously been traversed. It may also be the case, for some networks, that while required fact pre-condition values can be obtained for a given rule, they may not be obtainable at the same time, meaning that the precondition can never actually be satisfied and the vulnerability it represents should not be utilized in an attack design.

An efficient solution to this challenge is to define and store traversal paths in a Blackboard Architecture network and to also store, associated with them, the relevant container state information. Paths which consider the current state of containers, which can have multiple states, are called reality paths. Each state of each container that is involved in a reality path is stored in a simple structure called a container variant. The use of these reality paths and variants is described and analyzed herein and their efficacy is assessed.

This paper continues in Section 2 with a discussion of prior work which provides the foundation for the work presented herein. Then, in Section 3, an overview of the proposed system is presented. Next, in Section 4, the operations of the system are analyzed. Following this, Section 5 presents the experimentation and results that were collected. Then, Section 6 discusses these results and their implications, before the paper concludes, in Section 7, with a discussion of key conclusions and areas of potential future work.

## 2. Background

This section reviews prior work in several areas which provides a foundation for the work discussed herein. First, prior work on the Blackboard Architecture is reviewed. Then, prior techniques for concurrency management under the Blackboard Architecture are discussed.

### *2.1. Blackboard Architecture*

The Blackboard Architecture is based on a form of artificial intelligence called expert systems. Rule-fact expert systems are among the oldest forms of artificial intelligence. They originated in the 1960s and 1970s with Feigenbaum and Lederberg's Dendral system [1], which separated knowledge storage and processing [2] and Mycin (which is considered, by some, to actually be the first expert system) [2]. Classical expert systems implement a collection of facts which are interconnected by rules and can be used to perform inference [3]. Notably, this allows the implementation of both inductive and deductive forms of reasoning.

Expert systems, generally, are defined by their capability to state what they know and why they know that information to be true [4]. A variety of expert system enhancements have been proposed [5]. Examples of enhancements include implementing evolutionary genetics to optimize data [6] and adding support for fuzzy logic [7]. Expert system uses have included medical applications [8], education [9], robotics [10,11], engineering [12] and agriculture [13].

The Blackboard Architecture builds on this concept and was introduced by Hayes-Roth in 1985 [14], based on the Hearsay-II system [15,16]. The Blackboard Architecture builds upon the capabilities of expert systems, allowing them to actually make and take action based upon decisions, instead of just making recommendations. Thus, the Blackboard Architecture provides the capability to alter the system's operational environment. To implement this, the most basic form of a Blackboard Architecture system needs only to add action objects [17] to expert systems' rules and facts.

The Blackboard Architecture has been demonstrated for use in a variety of application areas including proof creation [18], handwriting recognition [19], robotics [20], production scheduling [21] and software testing [22].

### *2.2. Managing Concurrency in the Blackboard Architecture*

A number of approaches to concurrency management have been proposed for use with the Blackboard Architecture. Many of these are born from the development of distributed systems. One of the most basic approaches for creating a distributed blackboard system is was proposed by Compatangelo [23,24] and consisted of multiple agents which connected to a blackboard which was housed in shared memory. Building on this, Kerminen and Jokinen [25] developed a system with a centralized, non-replicated storage capability [26]. Redondo and Ortega [27], similarly, used a centralized storage server with tuples which could be checked out for use. Weiss and Stetter [28] used a hierarchical structure where lower-level systems communicated with higher level systems using "ambassador" nodes (which were based on work in [29]) that represented the lower-level systems. Adler [30], on the other hand, used a central system that provided limited storage and knew the location of remote data.

Jiang, et al. [31] (building on prior work by Botti, et al. [32]) proposed a system based on blackboard-to-blackboard communications – an approach which was also utilized by Jiang and Zhang [33]. Larner [34] considered a Blackboard Architecture system which might be distributed across multiple computers and discussed solutions for replication and concurrency. The approach may facilitate workload distribution; however, it might require highly distributed queries across numerous system nodes. A data replication approach was proposed by Saxena, et al. [35], where multiple nodes would store copies and replication would be conducted using multicast messaging. Velthuijsen, et al. [36], alternately, proposed the use of centrally managed locks to prevent concurrency problems. van Liere, et al. [37] suggested only synchronizing data that is actually used in a particular system, while a client-server model was utilized by Tait, et al. [38,39].

The use of boundary nodes to represent portions of a Blackboard Architecture network [40] has been demonstrated in prior work. Maintenance automation [41], which could facilitate concurrent independent operations, solving-based approaches [42,43], and pruning [44], which could reduce network search space and replication requirements, have also been previously proposed.

### 3. System Overview

This section provides an overview of the operations of the system. First, Section 3.1 provides an overview of the objects in the system. Then, Section 3.2 describes the operations of the system. In Section 3.3, reality paths are described. Next, in Section 3.4, variants are discussed. Finally, Section 3.5 presents the Sonar system.

### *3.1. System Objects*

The basic components of a Blackboard Architecture network are rules, facts and actions. Facts store binary values which are the information of the system. Rules are propositional logic connections between facts which assert an output fact if the required inputs are true and the rule is run. Actions are triggered by rules and used to actuate in the system's operational environment. These basic components allow many decision-making processes to be readily modeled.

Prior work also introduced a second form of organization, containers and links. Containers and links allow the physical, organizational or other similar structure of a phenomena to be modeled independently of decision-making logic; however, this structure can inherently be considered by the decision-making processes. Containers are collections of facts that relate to a single named entity. Links model relationships between entities.

Further building upon this, prior work also introduced common properties and generic rules. Common properties are facts that provide a specific labeled type of information regarding a container. A fact object is associated with a container and with a particular common property type, thus allowing facts that store the same information about different containers to be readily identified. Generic rules are rules that can act on any link-connected containers with the requisite common properties and required pre-condition values and can make changes to facts within both of the connected containers.

For modeling a computer network for security assessment purposes, containers are used to represent network devices (such as computers, servers, printers, routers and switches). Links are used to represent physical (including wireless) connectivity between devices.

Common properties were implemented for facts which could be utilized across a number of different containers. A simple example of this might be a fact called 'hasAdmin' which could indicate that a computer or device has a local admin account.

An object to model network pathways, called a reality path, is used to house information about paths of traversal in the Blackboard Architecture network. Each reality path stores updated fact information, for facts changed during a run iteration, in alternative container objects called variants. Reality paths and variants are discussed in more detail in Sections 3.3 and 3.4.

*3.2. System Operations*

An object-oriented design, presented in Figure 1, was used to implement variants and reality paths to collect the data which is presented and analyzed herein. Variants of containers and the facts created for them, which are separate instantiations based on the original objects and not data-bound to them, were used for the purpose of traversal.

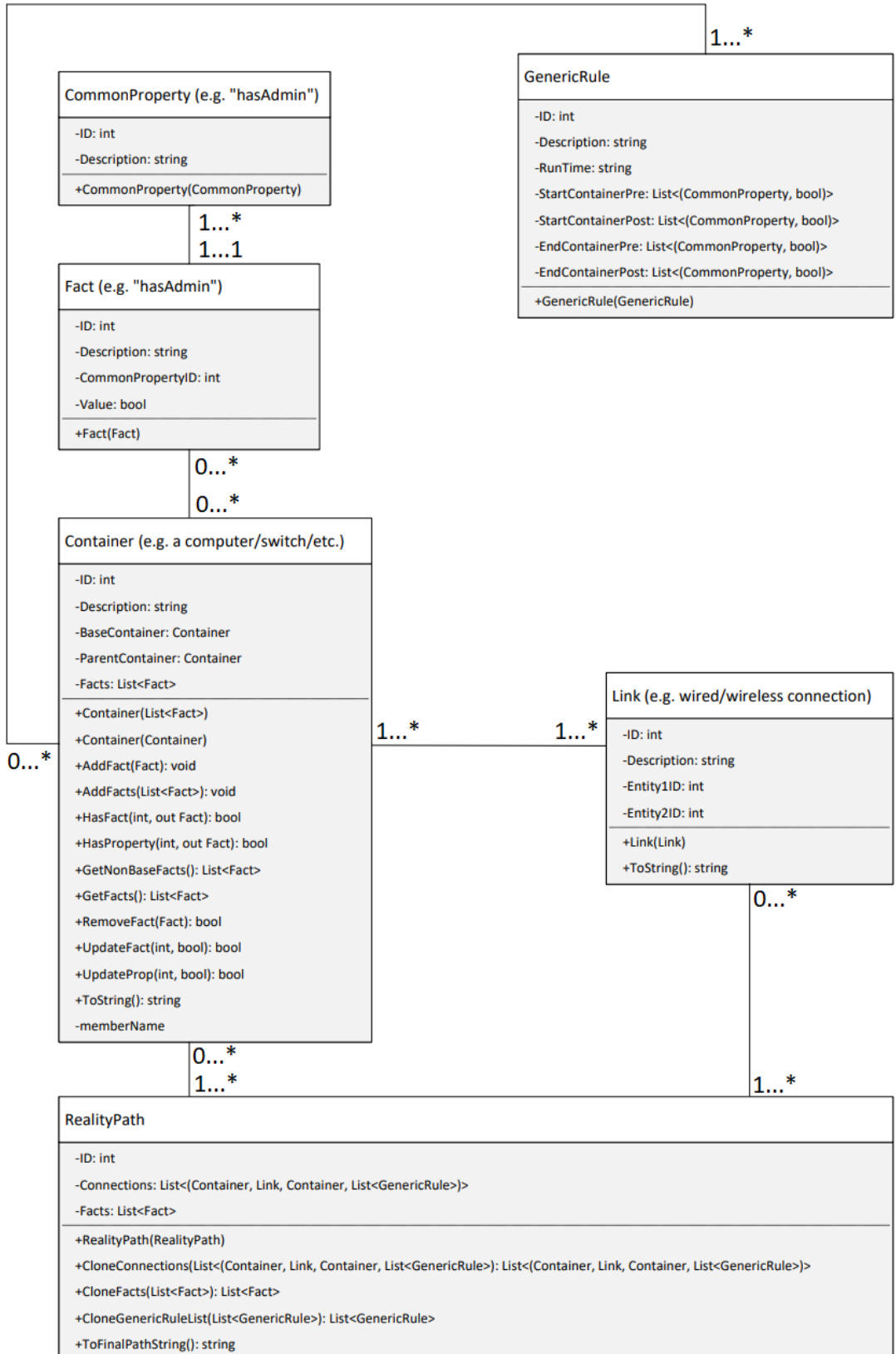

Figure 1. Unified modeling language (UML) diagram of the Blackboard Architecture including variants and reality paths.

Since facts are the only dynamic values in the system, at any given point in time the state of a given Blackboard Architecture network can be defined by which of its facts are true. In this system, facts do not stand alone and are always associated with containers. In the Blackboard Architecture implementation used for this work (discussed in section 3.5), containers are lists of fact objects with an identifying label.

Mathematically, containers can be thought of as having a relation which maps them to a set of facts. This relation can be expressed as $M_i(c_i) = \{f_1, f_2, f_3, ... f_n\}$, where each $c_i$ is in the set of containers, $C$, each fact, $f_1...f_n$, is in the set of facts, $F$, and $M_i$ is the mapping from the set of mappings, $M$, that is associated with container $c_i$. Rules are conditional statements where the antecedent and the conclusion are logical conjunctions of Boolean values in $F$. In other words, a rule can be defined as a function of the form $R = (a, b)$, where $a$ is a set of facts and the values which those facts must have for the rule to be run, and $b$ is a set of post conditions (facts and their values after the rule has run).

The set of all rules in the system is denoted as $R$. In the proposed implementation, rules do not change, and any rule or set of rules may be applied at any relevant point in time during traversal. Rules – since they can alter facts – have the potential to alter whether a given node is traversable or not.

To search for attack paths in a network modeled this way – where state-altering rules can be applied any number of times and at any point in time – one of several approaches could be taken, depending on the time and space requirements.

Two main approaches are discussed in this paper. Inductive approaches use methods of traversal which start from an ingress point and use heuristics, statistical modeling, or fixed constraints to search for most possible ways to reach the egress. It is possible that a given inductive approach might overlook potential attack paths in some cases. Deductive approaches are those which consider the set of all possible state transitions from the ingress to the egress and eliminate paths until only valid attack paths remain. A number of techniques could be employed to eliminate paths in this manner. For example, if a container is discovered during traversal which is only relevant in a known set of cases, it can be indexed (i.e., appended with data defining those cases in which it would be relevant) and removed from all further computations unless one of the cases should arise, thereby removing a set of potential paths from consideration in the continued exploration of the network.

A traversal algorithm, such as a depth-first or iterative deepening search, could be implemented deductively against a Blackboard Architecture system in such a manner that it did not terminate upon reaching a goal state but ran until all child states were closed. If the agent represented a threat actor, then an action by the threat actor would either be a move to control a different container, or a running of a rule to alter the state of the network. Thus, the agent would be searching the space $C \times F^*$ (the Cartesian product of the containers and the power set of facts in the system). Applying a depth-first queue-based algorithm would be computationally prohibitive for a few reasons. First, all true facts in the system would be indexed, even though most of them would not be relevant to the agent's current state, and this would unnecessarily increase the space complexity of the traversal algorithm. Second, for large networks with potentially thousands of global facts and/or rules which can be applied at arbitrary points in time, generating child states for each rule would result in thousands of children for each container in the network. In a worst-case network, in which each pair of containers is connected directly, and in which a rule or composition of rules exists to create every possible combination of facts at any given point in time, the number of possible paths could be computed as follows:

$$\text{Number of possible paths} = \sum_{k=0}^{k=(m\times 2^n)-2} {}_{[(m\times 2^n)-2]}P_k$$

where m is the number of containers in the network, and n is the number of facts in the network

### 3.3. Reality Paths

It is essential to track the transitions made while progressing through the network from the ingress to the egress; however, deductive approaches, which begin considering all possible paths, will typically be infeasible for networks beyond the smallest sizes. Thus, an alternative is required. Purely inductive approaches, such as those which rely solely on statistical modeling, would likely miss some attack paths.

Paths can be readily omitted unless all possible state transitions are considered to deduce which ones aren't necessary. For example, in a Blackboard Architecture network, traversal from reachable node *A* to unreachable node *B* may not be possible initially, but upon traversal to node *D* and the application of rules which were only accessible from *D* to the set of facts, the facts in the system altered in such a way that traversal from *A* to *B* becomes possible. In this example, had a simple algorithm been implemented at node *D*, node *A* would have been deemed closed after the initial traversal, and one or more attack paths would have been overlooked.

This presents an intriguing design dilemma. It is imperative to assess all of the transforms possible in the network, and this would need to include all information pertaining to the traversibility of the network. However, this must be done without being able to easily predict which rules will be of the most relevance. Problematically, assessing the entire state of the network for every transition would be computationally prohibitive.

The approach taken to this challenge was to create a data structure which could be used to represent a path through the network, and to reuse that same data structure across multiple fact configurations. Paths were stored as objects which modeled the containers and facts traversed through the network. The advantage with this design is that, if rules altered facts in the network, new states from the cartesian product *C x F\** don't need to be generated; instead, only the containers that were modified need to be mapped to the path. To generate children, all nodes/containers which could be reached either by immediate traversal or by an initial application of rule(s) before traversal are considered as states. In this case, only reachable containers in *C*, along with the fact configurations upon traversal, are considered children of a given state. This approach enabled considerable optimizations to be made to the traversal algorithm.

This study models paths formed in this way using a data structure called reality paths. Each reality path contains information about the containers which have been traversed, the order in which they were traversed, and the state of facts within them. It also includes information about what rules have been triggered and when, during this process. In this implementation, reality paths are implemented as .NET objects which house lists of containers which have been traversed and their state at the time of traversal.

### 3.4. Variants

To store a route through the Blackboard Architecture network, each reality path clones containers and their component facts which were altered during traversal. These are stored in lists in the reality path object, and another list was created to track the order that containers were traversed in and the rules applied. New paths are formed every time a new container is visited; however, the old paths still exist until they terminate if they are unable to generate child states.

The cloned facts and containers are called variants, and they serve as a means of storing the state changes within a given reality path of traversal. Thus, each reality path could have one or more variants for each container, as well as one or more variants for each given fact. When, during iteration, a container is reached, it is determined whether a variant of that container already exists in the given reality path or not. If it does, that variant is used. A list of container names is kept in the reality path object, storing the order of transitions.

The container variants are stored in memory so that they can be analyzed for future state generation. The system uses operating system pagination if physical memory limits are reached. There is no direct interface with the hard disk. Due to the number of possible paths and containers, optimizations were implemented.

The approach of cataloging traversals via reality paths presented additional considerations. First, it can result in traversal loops among containers in the network. To address this problem, a configurable cap was set as to the number of times a given link could be traversed in a particular reality path.

Another consideration is the potential difficulty of ensuring that critical paths aren't missed when a link cap is imposed or when pathing is limited to immediately traversible containers. This remains as an area in need of future study. Overall, reality paths and variants are crucial to the use of the Blackboard Architecture for a security assessment system. An example illustrates how these objects optimize the space complexity of a network.

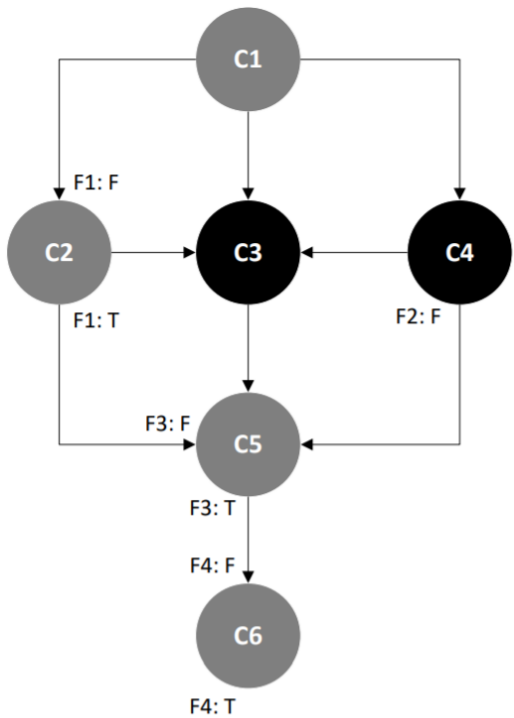 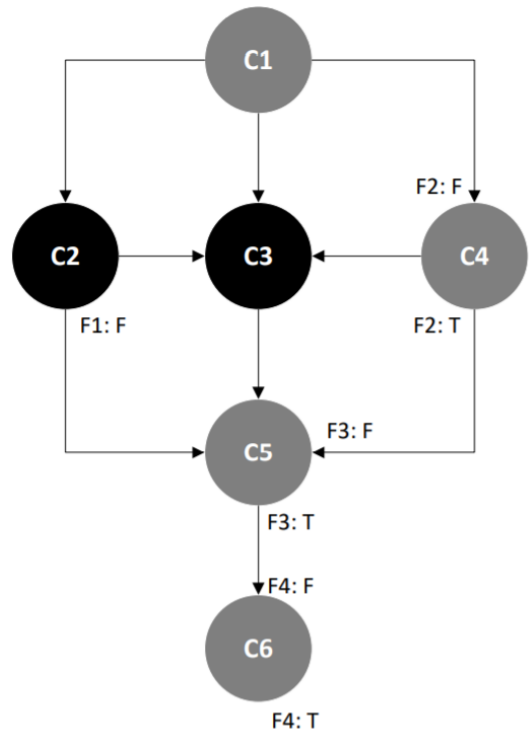

Figure 2 - Example Reality Path 1    Figure 3 - Example Reality Path 2

Figures 2 and 3 show how reality paths and variant containers are used for recording attack pathways through a given network. A starting node is provided and, as the network is traversed, different pathways with different variations of facts are generated.

Figures 2 and 3 show two example reality paths, among multiple possible paths, from container 1 to container 6. The two paths reach the end container with different fact configurations. The first reality path goes through container 2, container 5, and ends at container 6. This pathway would trigger a rule that alters fact 1 from false to true, and that fact alteration allows a rule on container 5 to alter fact 3 from false to true, ultimately allowing fact 4 in container 6 to change from false to true.

The containers that store these altered facts have become variant containers, which allow the reality paths to traverse nodes without affecting the original network, and other traversal by extension. The alterations within each reality path are specific to their path and do not affect other reality paths. The second example goes through container 4, container 5, and ends on container 6. Rules, for this path, result in fact 2 being altered from false to true in container 4. Fact 3 is altered from false to true in container 5, and fact 4 is altered from false to true in container 6. Notably, this second traversal is possible because rules reference the original network and the original facts within it, not the other reality path that has stored its fact alterations within variant containers. The end result is two different attack pathways containing unique attack information being identified within the network.

Without reality paths and variants, traversal would alter facts in the original network, preventing accurate traversals during simultaneous or subsequent runs. To fully explore all paths, the network would need to be reset back to the original state. With the network shown in Figures 2 and 3, the first reality path would have to be recorded and the altered facts would need to be set back to false. Then, the next run through the network would have to avoid the pathway that was previously found. This can be simplified by using reality paths to simultaneously traverse the network and record fact alterations to variant containers where they can be stored for further rule running within paths without interfering with the base network.

*3.5. SONARR*

A program called the software for operations and network attack results review (SONARR) was used to test the approach being proposed for traversal of Blackboard Architecture networks.

In SONARR, a list of paths is explored, and the children for each path are generated and added to the end of the list for further exploration. The program either terminates when a configurable cap to the number of nodes traversed is hit or, for on simpler networks, when no new nodes can be reached.

The set of children, for a given path, is computed as the union of two sets. The first is the set of all adjacent containers with no alterations in the fact space of the network (i.e., the Cartesian product of all adjacent containers and the current fact state). The second is all adjacent containers with each of the application rules that altered facts in the current and the adjacent container (in other words, the union of each set of adjacent containers composed with each one of the rules that applied to it and the current container). Child states are stored in variants; however, new variants are not generated if a matching variant already exists in a given reality path. New reality paths are generated when new containers are traversed to. SONARR presently only applies one rule per child generated, and does not consider compositions of rules. To prevent the algorithm from running in an unending loop, a configurable cap to the number of times links are traversed was included.

**4. System Analysis**

This section considers a mathematical analysis of system operations. Empirical data is analyzed in the following section.

Since facts can be either true or false, there exist $2^n$ possible combinations of facts for a given network of n facts. There are m container nodes, including the ingress and egress. Each of these could be visited while a distinct configuration of fact settings existed, if not considering the limitations posed by rules' state-change capabilities. Given this, there could be $(m)(2^n)$ possible states, making it so there are $(m)(2^n)-2$ possible states between the ingress and egress. Paths through the network are permutations or orderings of these states. The number of state transitions in a valid path can vary significantly, from one to the number of possible intermediate states. Summing the number of paths consisting of one state transition, the number of paths consisting of two state transitions, and so forth through the number of paths incorporating all states, gives the total number of possible paths in the network. Using the permutation formula, $_nP_r$, for each number of state transitions, the maximum number of paths can be computed with the formula:

$_{[(m)(2^n)-2]}P_1$ *permutations of network states which visit one intermediate state,*

$_{[(m)(2^n)-2]}P_2$ *of network states which visit two intermediate states,*

...

*and* $_{[(m)(2^n)-2]}P_k$ *permutations of network states which visit k intermediate states*

Summing all possible numbers of states from 1 to $(m)(2^n)-2$, the number of possible paths can be calculated using the equation:

$$Number\ of\ possible\ paths = \sum_{k=0}^{k=(m\times 2^n)-2} {}_{[(m\times 2^n)-2]}P_k$$

Depending on the size of the network being modeled, computing paths as state transitions within the state space could quickly become computationally unfeasible. Optimizations, thus, need to be implemented.

An algorithm could prospectively be achieved in two ways: either by starting with all possible paths through the network and pruning the invalid ones, or by starting with no paths through the network and adding the valid ones.

A number of deductive pruning-based optimizations were considered; however, these approaches proved to be impractical due to the size of the state spaces. Computations of their size are discussed in Appendix A. They prove to be very computationally intensive, even for small networks.

One approach that was evaluated was combining both deductive and inductive methodologies. This involved using a depth-first search algorithm to generate all possible paths through the state space. Then, each child state was stored, augmented with data indicating the specific conditions under which the children would be relevant to the goal. If those conditions are not present in the system, then the state could be regarded as closed. During program execution, the closed rules are periodically reviewed, and if their conditions are now satisfied, they are placed back into the traversal queue. Using this method, paths can be explored more rapidly since their children are not explored. This approach makes it possible to exclude large numbers of paths from consideration during traversal, expediting the process.

Bit packing techniques were used to store large amounts of fact data as collections of bits. Computed paths could be stored in a data structure that would be traversable using a binary search algorithm. Paths created using this data structure could be stored and retrieved from the hard disk in batches for efficiency.

The use of metadata tags or reverse-indexing (i.e., indexing states which aren't necessary to traverse or indexing specific conditions, which, if met, would retrieve closed state(s) for further processing) are approaches which may merit further investigation.

In this study, a variation of this concept was implemented. Since this approach stores fact transitions (in addition to containers), its paths could be stored as a set of the containers traversed, the facts which were true upon traversal, and the facts which are indexed (i.e., stored to indicate specific conditions under which the path becomes valid, for the sake of traversal optimization). These paths are of the form:

$P = \{(c_0,f_0), (c_1,f_1), (c_2,f_2), ...(c_n,f_n) : c_i \in C, f_i \in F\}$

where $f_i \in F$ is the set of facts which have changed upon traversal to the container and $c_i \in C$ is the rule used for the traversal.

Reality paths of this form – or any mathematical equivalent thereof – enable a well-designed algorithm to evaluate all possible routes through a network. However, traversing using this data structure using a brute force approach greatly increases the time complexity of the algorithm.

This approach was ultimately deemed unsuitable due to a number of factors. The most notable of which are the size of the state space and complexity of the algorithm. While it provides optimization benefit, it presents several issues. First, programs implementing this model are fact-agnostic and, thus, incapable of deciding which paths are most relevant. Several redundant paths must be generated for different fact configurations even in the best cases. This makes the resulting list of paths – even with optimization applied – very large. Additionally, a disk interface-based program would not be as versatile or extensible.

## 5. Experimentation and Results

To assess the functionality of reality paths and variants, five common network topologies (bus, mesh, ring, star, and tree) were modeled in SONARR and the results are compared. Many networks are a complex amalgamation of topologies, but the five listed topologies are common substructures in most networks, and they, thus, provide a basis of comparison in terms of typical patterns the proposed system would be likely to encounter.

Bus and mesh topologies are indistinguishable, from the perspective of a SONARR model, so the results for these topologies are discussed together. Twenty trials were run for each topology, and the results were averaged. These average values are presented. In some cases, when the space complexity of a given topology increased exponentially, the program was terminated after one hour of runtime, and the results are compared at that point.

The averages of the data are collected during experimentation are presented now. A diagram of the experimental ring topology is shown in Figure 4, and the data collected using it is presented in Table 1. The experimental tree topology is shown in Figure 5, and its data is presented in Table 2. The star topology and data are presented in Figure 6 and Table 3, respectively. Finally, the Bus and Mesh topologies are depicted in Figure 7, and their data is presented in Table 4.

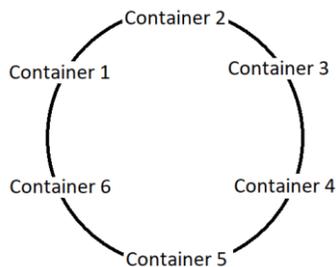

Figure 4. Ring topology (run from container 1 to container 4).

Table 1. Performance for ring topology.

| Link cap | Average Number of Reality Paths | Average Number of Variants | Average Run time |
|---|---|---|---|

| 1 | 6 | 6 | < 0.001 seconds |
| 2 | 190 | 170 | 0.017 seconds |
| 3 | 7,098 | 6,482 | 10.82 seconds |

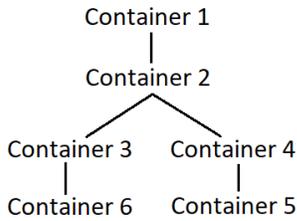

Figure 5. Tree topology (run from container 5 to container 6).

Table 2. Performance for tree topology.

| Link cap | Average Number of Reality Paths | Average Number of Variants | Average Run time |
|---|---|---|---|
| 1 | 2 | 4 | < 0.01 seconds |
| 2 | 96 | 192 | 0.017 seconds |
| 3 | 5642 | 11,284 | 28.12 seconds |

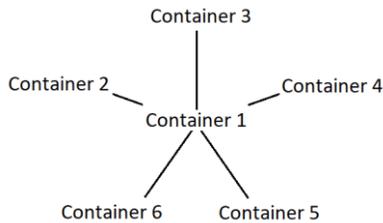

Figure 6. Star topology (run from container 1 to container 2).

Table 3. Performance for star topology.

| Link cap | Average Number of Reality Paths | Average Number of Variants | Average Run time |
|---|---|---|---|
| 1 | 65 | 130 | 0.004 seconds |
| 2 | 7,365 | 14,730 | 26.2 seconds |
| 3 (stopped at 1 hour) | 70,627 | 141,278 | 1 hr. (stopped) |

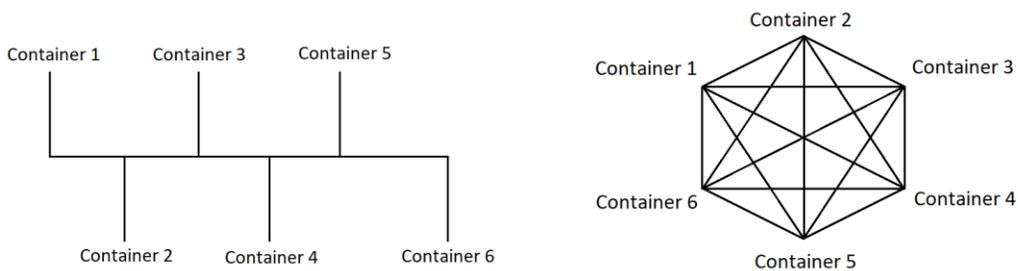

Figure 7. Bus or Mesh topology (run from container 1 to container 4).

Table 4. Performance for bus/mesh topology.

| Link cap | Average Number of Reality Paths | Average Number of Variants | Average Run time |
| --- | --- | --- | --- |
| 1 (stopped at 1 hour) | 435,239 | 49,218 | 1 hr. (stopped) |
| 2 (stopped at 1 hour) | 222,113 | 44,964 | 1 hr. (stopped) |
| 3 (stopped at 1 hour) | 156,072 | 44,449 | 1 hr. (stopped) |

From this data, it is clear that the complexity of the network is the most significant factor in the number of paths generated. More complex, highly interconnected networks will, thus, particularly require additional methods of optimization. The link cap, which limits the number of times a given link is traversed, has the potential to significantly reduce the number of paths, but it does this at the cost of potentially failing to identify some valid paths. Methods of optimization remain as an area where future study is required.

## 6. Evaluation and Discussion

This section discusses the data presented in Section 5 and draws conclusions regarding the performance of the proposed structures and system.

Inductive path discovery using metadata or object variants as a means of algorithmic optimization has the potential to dramatically reduce the number of paths identified as compared to the total number that are theoretically possible. In many cases, this reduction will be by many orders of magnitude. This is a primary benefit of the approach.

However, this approach has a key limitation of having the potential to enter potential loops. To prevent this, A limit parameter was added that caps the number of links traversed during computation. This dramatically reduces the number of results found; however, it risks the loss of complex attack paths requiring large attack loops.

Notably, there is an apparent trend where the more highly interconnected networks (tree and star networks), tend to produce higher ratios of variants to reality paths. These ratios tend to increase as the link cap is raised. It is hypothesized that this trend is due to the number of facts that a container has adjacent to it. In more highly-connected networks, agents operating from a container can traverse to several adjacent containers. Assuming those containers have more than one fact each, the number of facts the agent can traverse to, thus, increases at a higher rate than the number of containers it can traverse to. Thus, the number of variants increases more quickly in those cases. However, in simpler networks, an agent is limited in the number of directions it can traverse to and, thus, will increasingly witness common variants of facts recurring within the nearest containers. In these cases, especially where there is a higher link cap, traversal among the same containers will result in a comparitively higher number of moves through the network (since previously-computed variants are not traversed to), resulting in a higher ratio of paths to variants.

The data collected for the bus/mesh topology is limited in value, since SONARR was terminated prior to natural completion in these cases. Given this, it is possible that these data are not fully reflective of what the ratios of variants to paths would be if the program ran until completion. The primary value of this data, thus, comes from comparing system performance under this network topology to other topologies. The number of variants remained relatively consistent, while the number of paths was reduced as the link cap was increased. It is possible that this was due to a limited number of reachable fact configurations, which kept the variants within a narrow range. A higher link cap may have resulted in more moves throughout the network being amongst pre-existing variants rather than among new containers, which may explain the decreasing number of reality paths and the non-increasing number of variants.

Given the reductions in computed paths by using variants and link caps, it is apparent that an inductive approach carries sufficient obstacles; however, a purely deductive approach – since it would generate even more possible paths – would likely be nearly impossible, without significant computational complexity.

**7. Conclusions and Future Work**

The work presented herein shows that an inductive approach, which discovers paths rather than prune-optimizing from a collection of all possible paths in a Blackboard Architecture network, appears to be the preferred approach. The key limitation to this approach is that it may overlook possible valid attack paths. A clear trade-off exists between runtime and search completeness. Further optimizations may be possible through the use of heuristics and other techniques. Some of these may facilitate speed enhancements while having a reduced impact on search completeness.

Inductive approaches discover paths using information from the current state of the network, while deductive ones attempt to prune the set of all possible paths to only valid ones. Deductive approaches, which account for all possible state transitions through the network, must take into consideration the potential for rules to alter facts at arbitrary times, thus, altering the possible state transitions. In large networks, accounting for all possible state transitions during path computation becomes computationally infeasible.

Inductive path discovery poses its own limitations. Since rules may alter the system in ways which aren't immediately relevant to traversal, but which may become relevant later, it is difficult to predict which rules should be run and which ones can be disregarded. Because of this, purely inductive approaches which do not track all state changes at all times in the network may overlook possible attack paths.

A hybrid method analyzes the state changes which are of the most relevance to an agent in the network which is traversing from one container to another. The network variations which are caused by the traversal are stored in program memory, and child paths are branched from this state. These paths and variations are housed in objects called reality paths and variants.

This paper has shown that the use of the new reality path and variant structures can significantly reduce the amount of time required to search for potential cybersecurity vulnerability paths. While this approach is notably faster than deductive approaches, its completeness cannot be guaranteed while applying a link cap. Without link caps, at present, the system could end up in a perpetual loop.

Current implementations of reality paths and variants are also still computationally expensive. Further optimizations may be possible through a combination of indexing (i.e., grouping paths with data pertaining to conditions which would make the path relevant or not and using this data to retrieve or eliminate paths from consideration), which will reduce the set of paths considered, and the use of heuristics, which will enhance decision-making during path generation.

**Acknowledgements**

This work has been funded by the U.S. Missile Defense Agency (contract # HQ0860-22-C-6003).  Thanks are given to Jonathan Rivard and Jordan Milbrath for their work on the software system that this technology operates within.

**Appendix A. Approximation of the Number of Possible Paths**

An exhaustive search approach could start from the generation of every possible path and prune paths which are impossible (and potentially those that are infeasible or even of a lower priority). To do this, it would start by generating every possible path. As attackers can use attack steps to make changes on a node that they are occupying or a remote node and can also move their location, the number of possible states that the system can be in is a combination of the attacker's location and the system's facts' values. For a system with m containers (locations) and n rules, overall, the number of possible states would be:

$$Number\ of\ possible\ states = m \times 2^n$$

One of these states is always a defined ingress (starting) point and one is a defined egress (ending) point. As shown in Figure A1, the shortest possible path is directly from ingress to egress. There is only one possible path of this type. Progressively longer paths are also possible. The second shortest path length has a single intermediate state. In a network with three container nodes, which each have three facts, the number of possible states would be equal to $m \times 2^n = 3 \times 2^9 = 1536$. The ingress and egress are two of these states, so there are 1534 possible intermediate state options and, thus, for a single intermediate state path, 1534 possible paths.

There are numerous length options which range up to a point where all 1536 states are visited. This is depicted in Figure A1, with the bottom network showing how the number of possible options for intermediate states would start at 1534 and decline to 1, for the longest network of 1534 intermediate nodes.

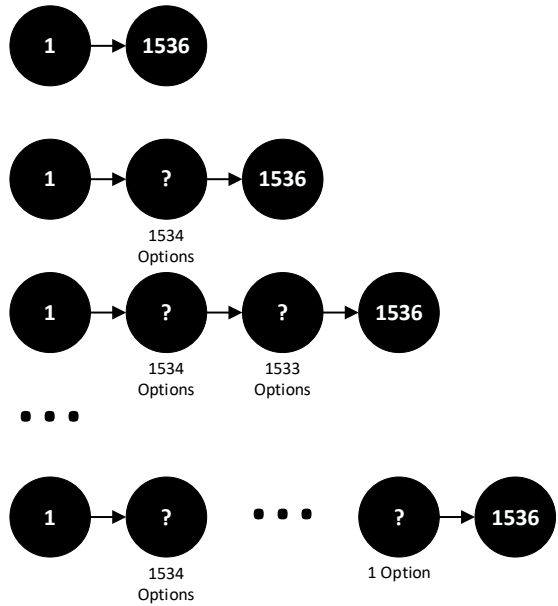

Figure A1. Possible paths for a 3 container, 9 fact network.

The number of possible networks of each length can be determined by multiplying the possible options together for each intermediate node. Thus, there would be 1534 single intermediate node paths and 1534 x 1533 = 2,351,622 paths with two intermediate nodes. The longest path length, shown in the bottom row of Figure A1, would have 1534! possible paths. The number of possible paths can be calculated with the equation:

$$Number\ of\ possible\ paths = \sum_{k=0}^{k=(m\times 2^n)-2} {}_{[(m\times 2^n)-2]}P_k$$

Because the paths that have less than the full chain of possible states remove the last multiplications, (e.g., one less than full removes 1!, two less than full removes 2!, three less than full removes 3!), the several largest chains will provide the most significant digits. By the point that 5! Is being removed from the end of the chain, the contribution is being divided by 120, effectively removing its contribution to the most significant digits. Thus, the first five longest paths can be used to approximate the entire value. Also, because these paths will always be the longest factorial divided by 5!, 4!, 3!, 2! and 1!, and added together, a coefficient can be calculated to apply. Thus, the number of possible paths can be approximated using the equation:

$$Number\ of\ possible\ paths \approx 1.71\overline{6} \times [(m \times 2^n) - 2]!$$

Using this for m = 3, n=9, thus, the number of possible paths is calculated as:

$$Number\ of\ possible\ paths \approx 1.71\overline{6} \times [(3 \times 2^9) - 2]!$$
$$Number\ of\ possible\ paths \approx 1.71\overline{6} \times 1534! = 1.19 \times 10^{4223}$$

From this large number, it is clear that this would be far too many paths to be processed within a feasible amount of time for most applications.